\documentclass[twocolumn,amsmath,aps]{revtex4}
\usepackage{graphicx,color}

\newcommand{\nl}{\nonumber \\}

\newcommand{\be}{\begin{equation}}
\newcommand{\ee}{\end{equation}}
\newcommand{\bea}{\begin{eqnarray}}
\newcommand{\eea}{\end{eqnarray}}

\newcommand{\Eq}[1]{Eq.\,(\ref{#1})}

\newcommand{\la}{\langle}
\newcommand{\ra}{\rangle}
\newcommand{\dg}{\dagger}

\begin{document}

\title{Weak Measurement of Qubit Oscillations
       with Strong Response Detectors:
       Violation of the Fundamental
       Bound Imposed on Linear Detectors}


%
%
%
%
%

\author{HuJun Jiao$^1$, Feng Li$^1$, Shi-Kuan Wang$^2$,
Xin-Qi Li$^{1,3}$ \footnote{E-mail: xqli@red.semi.ac.cn.}  }

\address{$^1$ State Key Laboratory for Superlattices and
Microstructures, Institute of Semiconductors, Chinese Academy of
Sciences, P.O.~Box 912, Beijing 100083, China}
\address{$^2$ Department of Chemistry, Beijing Normal University,
Beijing 100875, China }
\address{$^3$ Department of Physics, Beijing Normal University,
Beijing 100875, China }


\date{\today}

\begin{abstract}
We investigate the continuous weak measurement of
a solid-state qubit by single electron transistors
in nonlinear response regime.
It is found that the signal-to-noise ratio can violate
the universal upper bound imposed quantum mechanically
on any linear response detectors.
We understand the violation by means of the cross-correlation
of the detector currents.
\end{abstract}

\maketitle


Single electron transistor (SET) is a sensitive charge-state
detector \cite{Dev00,Sch01,Rim03}, which promises the use
for fast qubit read-out in solid-state quantum computation.
For \emph{single-shot} measurement, i.e., in one run
the qubit state is unambiguously determined,
an important figure of merit is the detector's efficiency,
defined as the ratio of information gained time
and the measurement induced dephasing time \cite{Sch01}.
In the weakly responding regime, it was found that SET has
rather poor quantum efficiency \cite{Sch01,Kor01,Moz04}.
However, recent study showed that, for strong response SET,
the quantum limit of an ideal detector can be reached,
resulting in an almost pure conditioned state \cite{Wis06}.

Rather than the single-shot measurement, a more implementable
approach in experiment is the continuous weak measurement.
This type of measurement allows the ensemble average of
detector and qubit states,
and the qubit coherent oscillation is read out from the
spectral density of the detector.
In continuous weak measurement, a remarkable result is the
Korotkov-Averin (K-A) bound, originally stated as follows \cite{K-A01}:
{\it The interplay between the information acquisition
and the backaction dephasing of the oscillations
by the detector imposes a fundamental limit, equal to four, on
the signal-to-noise ratio of the measurement.
The limit is universal, e.g., independent of the coupling strength
between the detector and system,
and results from the tendency of quantum measurement to localize the system
in one of the measured eigenstates }.
In order to overcome the K-A bound, special techniques
such as the quantum nondemolition (QND) measurement \cite{Ave02},
the quantum feedback control \cite{Wang07},
and the measurement with two detectors \cite{But06}, have been proposed.
In this work we investigate the continuous weak
measurement by strongly responding SETs \cite{Wis06,Gur05}.
Remarkably, we find that
for both models studied in Refs.\ \onlinecite{Wis06}
and \onlinecite{Gur05} the signal-to-noise ratio (SNR)
can violate the \emph{universal} Korotkov-Averin bound.
Interpretation and implication of this new result will be also provided.

\begin{figure}[h]
\begin{center}
\includegraphics[width=10cm]{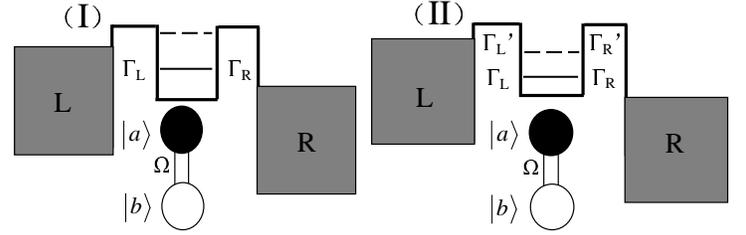}
\caption{\label{fig1} Schematic model for a solid-state qubit
measurement by SET. Model (I): the SET dot level is within the bias
voltage for qubit state $|b\ra$, but outside of it for state
$|a\ra$. Model (II): the SET dot level is between the Fermi levels
for either $|b\ra$ or  $|a\ra$, but with different coupling
strengths to the leads, i.e., $\Gamma_{L/R}$ for $|b\ra$, and
$\Gamma'_{L/R}$ for $|a\ra$. }
\end{center}
\end{figure}



{\it Model}.---
Consider a charge qubit, say, an electron in a pair of coupled
quantum dots, measured by a single electron transistor,
as schematically shown in Fig.\ 1.
The entire system is described by the following Hamiltonian
\begin{subequations}
\begin{align}
H &= H_0+H^\prime\\
H_0 &= H_S+\sum_{\lambda=L,R}\epsilon_{\lambda k}d_{\lambda k}^\dag
d_{\lambda k}   \label{Eq1}\\
H_S &= \sum_{j=a,b}E_j|j\ra\la j|+\Omega(|a\ra\la b|+|b\ra\la a|)
 + E_c a_c^\dag a_c +Un_an_c\\
H^\prime &=\sum_{\lambda=L,R;k}(\Omega_{\lambda k}a_c^\dag d_{\lambda
k}+{\rm H.c.}) \equiv ~ a_c^\dag (f_{cL}+f_{cR})+ {\rm H.c.}.
\end{align}
\end{subequations}
For simplicity, we assumed spinless electrons.
The system Hamiltonian, $H_S$, contains qubit and the SET central dot,
and their Coulomb interaction (the $U$-term).
For qubit, we assumed that each dot has only one bound state,
i.e., the logic states $|a\ra$ and $|b\ra$ with energies $E_a$
and $E_b$, and with a coupling amplitude $\Omega$.
$n_a$ is the number operator of qubit state $|a\ra$, which
equals 1 for $|a\ra$ occupied and 0 otherwise.
For the SET, $a_c^\dag (a_c)$ and $d_{\alpha k}^\dag (d_{\alpha k})$ are
the electron creation (annihilation) operators of the central dot and
reservoirs. $n_c\equiv a_c^\dag a_c$ is introduced as the number operator
of the SET dot. Similar to previous work,
we assumed that the SET works in the strong Coulomb-blockade regime,
with only a single level $E_c$ involved in the measurement process.
Finally, $H'$ describes the tunnel coupling of the SET dot to the leads,
with amplitudes $\Omega_{\lambda k}$.

In this work, we consider two SET models as schematically
shown in Fig.\ 1.
In model (I), which was studied in Ref.\ \onlinecite{Gur05},
the SET dot level is within the bias voltage
if the qubit is in state $|b\ra$, but it locates above the Fermi levels
when the qubit state is switched to $|a\ra$.
For state $|b\ra$, a nonzero current $I_b$ flows through
the SET; however, for state $|a\ra$, the SET current $I_a$ is zero.
Then, the qubit state can be discriminated from these different currents.
In this model, the signal current $\Delta I\equiv |I_b-I_a|$ is
twice the average current $\bar{I}\equiv (I_b+I_a)/2$.
In this sense, it is not a weak response detector.
In model (II), which allows to
illustrate the crossover from weak to strong responses,
the SET dot level is always between the
Fermi levels of the two leads, for qubit either in state
$|b\ra$ or in state $|a\ra$, but with different coupling strengths
to the leads, i.e., $\Gamma_{L(R)}$ and $\Gamma'_{L(R)}$.
For the convenience of description, we further
parameterize the tunnel couplings as
$\Gamma_L(\Gamma_L^\prime)=(1\pm\xi)\bar{\Gamma}_L$,
$\Gamma_R(\Gamma_R^\prime)=(1\pm\zeta)\bar{\Gamma}_R$,
and $\gamma=\bar{\Gamma}_R/\bar{\Gamma}_L$.
Here, $\bar{\Gamma}_{L(R)}=(\Gamma_{L(R)}+\Gamma'_{L(R)})/2$
denote the average couplings,
while $\xi$ and $\zeta$ characterize the response strength of the
detector to qubit. In this context,
we would like to stress that most previous works
were largely restricted in the weak response regime by assuming
$\xi\ll 1$ and $\zeta\ll 1$, except in Ref.\ \onlinecite{Wis06}
the quantum efficiency was investigated
in strong response regime using this model.

{\it Method}.---
In continuous weak measurement, the detector's output is
characterized by the current and noise spectral density.
For their calculation, the most efficient approach is the particle-number
resolved master equation \cite{Sch01}.
In obtaining it, the qubit and SET dot are regarded as the system of interest,
while the two leads of the SET as an environment;
and the tunnel coupling $H'$ of SET is treated perturbatively as an
interaction between them.
Up to the dominant second-order of $H'$, following
Ref.\ \onlinecite{Li051} we have
\begin{align}\label{Eq2}
 \dot{\rho}^{(n_R)}=&-i\mathcal{L}\rho^{(n_R)}-\frac{1}{2}\{[a_c^\dagger,
 A_{cL}^{(-)}\rho^{(n_R)}-\rho^{(n_R)}A_{cL}^{(+)}]\nl&
 +a_c^\dagger A_{cR}^{(-)}\rho^{(n_R)}
 +\rho^{(n_R)}A_{cR}^{(+)}a_c^\dagger
 \nl&-[a_c^\dagger\rho^{(n_R+1)}A_{cR}^{(+)}+A_{cR}^{(-)}
 \rho^{(n_R-1)}a_c^\dagger]+ {\rm H.c.}  \} .
\end{align}
$\rho^{(n_R)}$ is the reduced density operator of the {\it system}
conditioned on the electron number ``$n_R$" tunnelled through
the right junction (similar equation holds also for the left junction).
For simplicity, throughout this paper we use the convention
$\hbar=e=k_B=1$.
In \Eq{Eq2} the Liouvillian $\mathcal{L}$ is defined by
$\mathcal{L}(\cdots)=[H_S,\cdots]$, and the operators
$A_{c\lambda}^{(\pm)}\equiv C_{\lambda}^{(\pm)}(\pm\mathcal{L})a_c$.
The superoperoters $C_{\lambda}^{(\pm)}(\pm\mathcal{L})$
are the generalized spectral functions:
$C_{\lambda}^{(\pm)}(\pm\mathcal{L})=\int_{-\infty}^{+\infty}dt
C_{\lambda}^{(\pm)}(t)e^{\pm i\mathcal{L}t}$,
where the bath correlation functions
$C_{\lambda}^{(+)}(t)=\langle f_{c\lambda}^{\dg}(t)f_{c\lambda}\rangle_B$
and $C_{\lambda}^{(-)}(t)=\langle
f_{c\lambda}(t)f_{c\lambda}^{\dg}\rangle_B$,
and the average $\la \cdots \ra_B\equiv {\rm Tr}_B[(\cdots)\rho_B]$, with
$\rho_B$ the local thermal equilibrium state of the SET leads
determined by the respective chemical potentials.

Rich information is contained in the above particle-number resolved
master equation, since the conditional density matrix $\rho^{(n_R)}(t)$
is directly related to the distribution function,
$P(n_R,t)={\rm Tr}[\rho^{(n_R)}(t)]$,
where the trace is over the system states.
For instance, by virtue of this relation, the measurement current
can be obtained as \cite{Li051}:
\begin{align}\label{IRt}
 I_R(t) = \sum_{n_R}{\rm Tr} \{n_R \dot{\rho}^{(n_R)}(t)\}={\rm Re}{\rm Tr} [(a_c^\dagger  A_{cR}^{(-)}
-A_{cR}^{(+)}a_c^\dagger) \rho(t)],
\end{align}
where $\rho(t)\equiv\sum_{n_R} \rho^{(n_R)}(t)$ satisfies the
usual {\it unconditional} master equation,
by summing the above \Eq{Eq2} over $n_R$.

In continuous weak measurement, the detector's power spectral density
contains very useful information of qubit's coherent oscillation.
Formally, the noise spectrum of current consists of three
terms \cite{Li053}:
$ S(\omega)=\alpha S_L(\omega)+\beta S_R(\omega)
-\alpha\beta\omega^2S_N(\omega)$,
with $S_{L/R}(\omega)$ the noise of the left (right)
junction current $I_{L/R}(t)$,
and $S_N(\omega)$ the fluctuations of the electron number $N(t)$
on the central dot of SET.
$\alpha$ and $\beta$ are two coefficients
determined by the junction capacitances \cite{Li053},
and satisfy $\alpha+\beta=1$.
Further, for $S_{L/R}(\omega)$, it follows the MacDonald's formula
\begin{align}\label{SW-1}
S_{\lambda}(\omega)=2\omega\int_0^{\infty}dt {\rm sin}\omega t
\frac{d}{dt}[\langle n_{\lambda}^2(t)\rangle-(\bar{I}t)^2],
\end{align}
where $\bar{I}$ is the stationary current and $\langle
n_{\lambda}^2(t)\rangle$ = $\Sigma_{n_{\lambda}}n_{\lambda}^2 {\rm
Tr}\rho^{(n_{\lambda})}(t)$ = $\Sigma_{n_{\lambda}} n_{\lambda}^2
P(n_{\lambda},t)$. 
In practice, instead of directly solving $P(n_{\lambda},t)$, the
reduced quantity $\la n_{\lambda}^2(t)\ra$ can be obtained more
easily by constructing its equation of motion \cite{Li051}, based on
the particle-number resolved master equation \Eq{Eq2}. Thus, we can
obtain
\begin{align}
\frac{d}{dt}\langle n_{\lambda}^2(t)\rangle = {\rm Tr}[2
\mathcal{J}^{(-)}_{\lambda}Q_{\lambda}(t)+
\mathcal{J}^{(+)}_{\lambda}\rho^{st}],
\end{align}
where the particle-number matrix reads $Q_{\lambda}(t)\equiv
\sum_{n_{\lambda}} n_{\lambda} \rho^{(n_{\lambda})}(t)$ and
$\rho^{st}$ the stationary density matrix. Here the superoperator
means
\begin{align}
\mathcal{J}^{(\pm)}_{\lambda}(\cdots)=&\frac{1}{2}[
A_{c\lambda}^{(-)} (\cdots) a_c^{\dagger} \pm a_c^{\dagger} (\cdots)
A_{c\lambda}^{(+)}\nl & +a_c (\cdots) A_{c\lambda}^{(-)\dagger} \pm
A_{c\lambda}^{(+)\dagger} (\cdots) a_c].
\end{align}
Inserting Eq.(5) into Eq.(4) and transforming into the frequency
domain, we obtain
\begin{align}
S_{\lambda}(\omega)=4 \omega {\rm Im}\{{\rm
Tr[\mathcal{J}_{\lambda}^{(-)} \widetilde{Q}_{\lambda}(\omega)]}\}
+2 {\rm Tr}\mathcal{J}_{\lambda}^{(+)} \rho^{st}-8 \pi \bar{I}^2
\delta(\omega),
\end{align}
in which $\widetilde{Q}_{\lambda}(\omega)=\int^{\infty}_{0}
Q_{\lambda}(t) e^{i \omega t}$. we can easily calculate
$\widetilde{Q}_{\lambda}(\omega)$ by solving a set of algebraic
equations after Fourier transforming the equation of motion of
$Q_{\lambda}(t)$, as have been clearly described in
Ref.\onlinecite{Li053}. For $S_N(\omega)$, which is the Fourier
transform of the correlation function $\langle
\{N(t),N(0)\}\rangle$, following Ref.\ \onlinecite{Li053} the
quantum regression theorem gives $ S_N(\omega)= 2 {\rm Re
Tr\{N[\tilde{\sigma}(\omega) +\tilde{\sigma}(-\omega)]\}}$.
$\tilde{\sigma}(\omega)$ is introduced as the Laplace transform of
$\sigma(t)\equiv {\rm Tr}_B [U(t) N \rho^{st} \rho_B U^{\dg} (t)]$,
where $U(t)=e^{-iH_S t}$.
Obviously, $\sigma(t)$ satisfies the same equation of the reduced
density matrix $\rho(t)$. The only difference is the initial
condition, for $\sigma(t)$ which is $\sigma(0)=N\rho^{st}$.


{\it Results}.---
For both models in Fig.\ 1, the states involved
are  $|1\rangle\equiv|0a\rangle$,
$|2\rangle\equiv|0b\rangle$, $|3\rangle\equiv|1a\rangle$,
and $|4\rangle\equiv|1b\rangle$.
In this notation $|0(1)a(b)\rangle$ means that the SET dot
is empty (occupied) and the qubit is in state $|a(b)\rangle$.
Applying \Eq{Eq2} to  model (I) results in
\begin{subequations}
\begin{align}
\dot{\rho}^{(n_R)}_{11}=&i\Omega[\rho^{(n_R)}_{12}-\rho^{(n_R)}_{21}]
+\Gamma_L\rho^{(n_R)}_{33}+\Gamma_R\rho^{(n_R-1)}_{33}\\
\dot{\rho}^{(n_R)}_{22}=&i\Omega[\rho^{(n_R)}_{21}-\rho^{(n_R)}_{12}]
-\Gamma_L\rho^{(n_R)}_{22}+\Gamma_R \rho^{(n_R-1)}_{44}\\
\dot{\rho}^{(n_R)}_{12}=&-i\epsilon\rho^{(n_R)}_{12}+i\Omega[\rho^{(n_R)}_{11}
-\rho^{(n_R)}_{22}] -\frac{\Gamma_L}{2}\rho^{(n_R)}_{12}\nl
&+\frac{\Gamma_L}{2}\rho^{(n_R)}_{34}+\Gamma_R\rho^{(n_R-1)}_{34}\\
\dot{\rho}^{(n_R)}_{33}=&i\Omega[\rho^{(n_R)}_{34}-\rho^{(n_R)}_{43}]
-(\Gamma_R+\Gamma_L)\rho^{(n_R)}_{33}\\
\dot{\rho}^{(n_R)}_{44}=&i\Omega[\rho^{(n_R)}_{43}-\rho^{(n_R)}_{34}]
+\Gamma_L\rho^{(n_R)}_{22}-\Gamma_R \rho^{(n_R)}_{44}\\
\dot{\rho}^{(n_R)}_{34}=&-i(\epsilon+U)\rho^{(n_R)}_{34}+i\Omega
[\rho^{(n_R)}_{33}-\rho^{(n_R)}_{44}]\nl &+\frac{\Gamma_L}{2}
\rho^{(n_R)}_{12} - (\Gamma_R+\frac{\Gamma_L}{2}) \rho^{(n_R)}_{34}
\end{align}
\end{subequations}
Here, $\epsilon=E_a-E_b$ and $\Gamma_{L/R}=2\pi|\Omega_{L/R}|^2g_{L/R}$,
with $g_{L/R}$ the density of states of the SET leads.
For simplicity, the assumption of wide-band limit implies
$\Omega_{L/R}\equiv\Omega_{L/R k}$,
and makes $\Gamma_{L/R}$ energy independent.
Also, low temperature and $U\gg \Omega$ were assumed to further
simplify the equations.
Similarly, for model (II), we have
\begin{subequations}
\begin{align}
\dot{\rho}^{(n_R)}_{11}=&i\Omega[\rho^{(n_R)}_{12}-\rho^{(n_R)}_{21}]
-\Gamma_L^\prime\rho^{(n_R)}_{11}+\Gamma_R^\prime\rho^{(n_R-1)}_{33}\\
\dot{\rho}^{(n_R)}_{22}=&i\Omega[\rho^{(n_R)}_{21}-\rho^{(n_R)}_{12}]
-\Gamma_L\rho^{(n_R)}_{22}+\Gamma_R \rho^{(n_R-1)}_{44}
\\
\dot{\rho}^{(n_R)}_{12}=&-i\epsilon\rho^{(n_R)}_{12}+i\Omega[\rho^{(n_R)}_{11}
-\rho^{(n_R)}_{22}]\nl
&-\frac{\Gamma_L+\Gamma_L^\prime}{2}\rho^{(n_R)}_{12}+
\frac{\Gamma_R+\Gamma_R^\prime}{2}\rho^{(n_R-1)}_{34}\\
\dot{\rho}^{(n_R)}_{33}=&i\Omega[\rho^{(n_R)}_{34}-\rho^{(n_R)}_{43}]
+\Gamma_L^\prime\rho^{(n_R)}_{11}-\Gamma_R^\prime\rho^{(n_R)}_{33}\\
\dot{\rho}^{(n_R)}_{44}=&i\Omega[\rho^{(n_R)}_{43}-\rho^{(n_R)}_{34}]
+\Gamma_L\rho^{(n_R)}_{22}-\Gamma_R \rho^{(n_R)}_{44}\\
\dot{\rho}^{(n_R)}_{34}=&-i(\epsilon+U)\rho^{(n_R)}_{34}+i\Omega
[\rho^{(n_R)}_{33}-\rho^{(n_R)}_{44}]\nl
&+\frac{\Gamma_L+\Gamma_L^\prime}{2}\rho^{(n_R)}_{12}-
\frac{\Gamma_R+\Gamma_R^\prime}{2}\rho^{(n_R)}_{34}
\end{align}
\end{subequations}
Except for the conditions leading to model (II),
other parameters are the same as above.

\begin{figure}[h]
\begin{center}
\includegraphics[width=8cm]{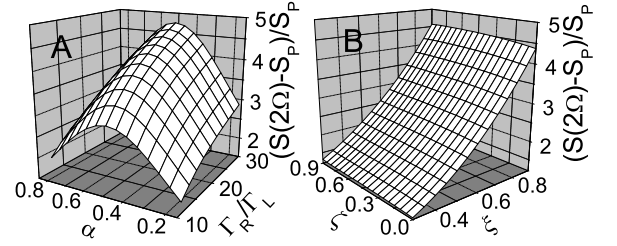}
\caption{\label{fig3} Signal-to-noise ratio: (A) for model (I), and
(B) for model (II).
For model (I), we take $\Gamma_L\equiv\Gamma$ as the energy unit,
and assume that $\mu_{L(R)}=\pm 50\Gamma$, $\Omega=2\Gamma$ and
$U=80\Gamma$.
For model (II), we use $\bar{\Gamma}_L\equiv\bar{\Gamma}$ as the
energy unit, and assume that $\Omega=\bar{\Gamma}$,
$U=50\bar{\Gamma}$, $\bar{\Gamma}_R=30\bar{\Gamma}$, and
$\alpha=\beta=1/2$. Also, zero temperature and $E_a=E_b$ are
assumed. }
\end{center}
\end{figure}

In continuous weak measurement of qubit oscillation,
the signal is manifested as a peak in the noise spectrum
at the qubit oscillation frequency $2\Omega$,
while the measurement effectiveness
is characterized by the signal-to-noise ratio (SNR),
i.e., the {\it peak-to-pedestal} ratio.
We denote the noise pedestal by $S_p$, and obtain it
conventionally from $S(\omega\rightarrow \infty)$.
In Fig.\ 2 we show the dependence of the SNR
on detector's configuration symmetries.

The result of model (I) is shown in Fig.\ 2(A), where
we see that both the tunnel- and capacitive-coupling
symmetries crucially affect the measurement effectiveness.
For the effect of tunnel coupling asymmetry $\Gamma_R/\Gamma_L$,
the basic reason is that,
with the increase of $\Gamma_R/\Gamma_L$, the interaction time
of the detector electron with the qubit is decreased.
Thus the detector's back-action is reduced and the SNR is enhanced
\cite{Gur05}.
For the effect of capacitive coupling, its degree of asymmetry
affects the contribution weight of the cross-correlation between
$I_L(t)$ and $I_R(t)$ to the entire circuit noise.
Specifically, the cross-correlation has more important contribution for more
symmetric coupling, as shown in Fig.\ 2(A) by the $\alpha$-dependence.
This is because, as we shall demonstrate below,
the cross-correlation has much higher {\it peak-to-pedestal} ratio
than the auto-correlation.

An unexpected feature observed in Fig.\ 2(A) is that under proper conditions,
say, the symmetric capacitive coupling and strongly asymmetric tunnel
coupling, the SNR can exceed ``4",
which is the upper bound quantum mechanically limited
on \emph{any linear response detectors} \cite{K-A01}.
However, to our knowledge, whether this upper bound is applicable
to \emph{nonlinear} response detector is so far unclear {\it in priori},
since in this case the linear response relation between current
and qubit state breaks down,
then the subsequent Cauchy-Schwartz-inequality based argument
leading to the upper bound ``4" does not work \cite{But06}.


To support the above reasoning, we further check model (II).
The result is presented in Fig.\ 2(B).
As explained in the model description, the parameters
$\xi$ and $\zeta$ used here characterize, respectively,
the left and right tunnel-coupling responses to the qubit states.
Shown in Fig.\ 2(B) is for an asymmetric tunnel coupling detector,
with $\gamma\equiv\bar{\Gamma}_R/\bar{\Gamma}_L=30$, which can
lead to higher SNR, because of the
weaker back-action from the detector, similar to model (I).
Here we find that the SNR is insensitive to the right
junction response $\zeta$, but sensitive to the left one $\xi$.
Again, in this model, we observe that the SNR can violate the
K-A bound ``4" in the strong response regime.

\begin{figure}[h]
\begin{center}
\includegraphics[width=8cm]{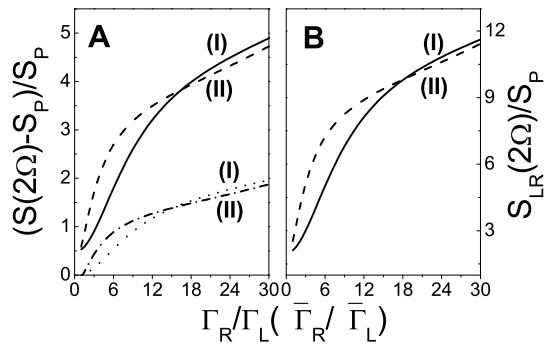}
\caption{Signal-to-noise ratio \emph{versus} tunnel-coupling
asymmetry, $\Gamma_R/\Gamma_L$ for model (I), and
$\bar{\Gamma}_R/\bar{\Gamma}_L$ for model (II). 
In (A) the solid and dashed lines are the result in the presence of
cross correlation, while the dotted and dot-dashed lines are the
result after removing it. In (B) the mere cross correlation is
plotted. 
$S_p$ is the pedestal noise of the entire circuit current.
$\xi=\zeta=0.9$, other parameters are the same as in Fig.\ 2. }
\end{center}
\end{figure}%

{\it Understanding the Violation of the K-A Bound}.---
Since $I(t)=\alpha I_L(t)+\beta I_R(t)$,
the current correlator $\la I(t) I(0) \ra$ contains the component
$S_{LR}(t)\equiv \la I_L(t) I_R(0)+I_R(t) I_L(0)\ra$,
i.e., the cross-correlation.
Also, in the previous results, we see that for
more symmetric capacitive coupling the SNR is larger, and
reaches the maximum at $\alpha=\beta=1/2$.
This feature indicates that the cross-correlation has an effect of
enhancing the SNR.
Indeed, for the SET detector, both the left and right junction currents
($I_L$ and $I_R$) contain the information of qubit state, so their
``signal" parts are correlated.
This leads to a heuristic opinion that views the two junctions
as two detectors, like the scheme of qubit measurement by two
point-contacts proposed recently by Jordan and B\"uttiker \cite{But06},
where they found that the SNR of the cross correlation can
strongly violate the K-A bound,
because of the negligibly small pedestal of the cross noise.
In our case, since $I_L(t)$ and $I_R(t)$ are subject to
a constraint from charge conservation,
the cross noise background of $I_L(t)$ and $I_R(t)$ does not vanish
in principle, unlike the two independent QPC detectors \cite{But06}.
Nevertheless, the pedestal of the cross noise of SET is much
smaller than that of the auto-correlation, which leads to an
enhanced SNR in the spectral density of the total circuit current,
and to the violation of the K-A bound, as clearly shown in Fig.\ 3(A).
For comparative purpose, in Fig.\ 3(B) we plot the SNR of the cross
correlation, scaled by the noise pedestal $S_p$ of the circuit current.

In Fig.\ 4 the spectral density of the cross correlation, scaled
by its own noise pedestal, is shown representatively.
As mentioned above, since at high frequency limit the cross noise pedestal
is negligibly small, here we artificially (but more physically in some sense)
define the pedestal at a finite frequency, e.g.,
twice the qubit oscillation frequency. Obviously, the giant SNR of the
cross-correlation has drastically violated the K-A bound.
This result indicates that in qubit measurement by SET
one can exploit the cross correlation, rather than the auto one as usual,
to probe the coherent oscillations. In practice, such scheme is simpler
than the technique of QND measurement \cite{Ave02},
and holds the most advantages of SET over QPC.
In recent years, the cross-correlation in mesoscopic transport
is an extensive research subject.
Its measurement in experiment is also possible, for instance,
using the nearby-QPC counting technique \cite{Ens06}.

\begin{figure}[h]
\begin{center}
\includegraphics[width=8cm]{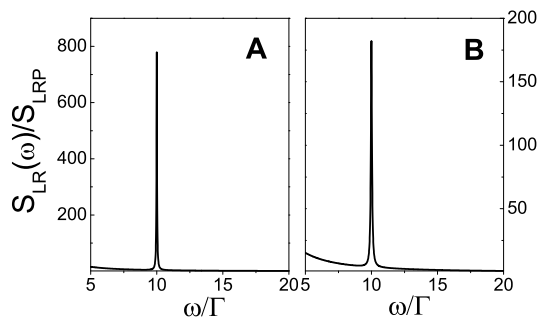}
\caption{ Spectral density of the cross-correlation scaled by its
own pedestal, here which is defined at twice the Rabi frequency of
the qubit oscillations. Parameters for model (I) in (A):
$\Gamma_L=0.05\Gamma$, $\Gamma_R=0.5\Gamma$, and $\Omega=5\Gamma$.
Parameters for model (II) in (B): $\bar{\Gamma}_L=0.05\Gamma$,
$\bar{\Gamma}_R=0.5\Gamma$, $\Omega=5\Gamma$, and $\xi=\zeta=0.9$.
$\Gamma$ in this figure is used as an energy unit, other conventions
are the same as in Fig.\ 2. }
\end{center}
\end{figure}

{\it Summary}.---
We have investigated the continuous weak measurement
of qubit oscillations by nonlinear response SET,
and demonstrated that the signal-to-noise ratio
can violate the universal Korotkov-Averin bound.
The violation has been understood by the role of the
cross-correlation of the detector's currents.
This interesting interpretation also leads to useful
implication to experiment.

{\it Acknowledgments}.
This work was supported by the National
Natural Science Foundation of China under grants No.\ 60425412 and
No.\ 90503013, the Major State Basic Research Project under grant
No.2006CB921201.

\begin{references}
\bibitem{Dev00}
M. H. Devoret and R. J. Schoelkopf, Nature (London) \textbf{406}, 1039 (2000).
\bibitem{Sch01}
Yu. Makhlin, G. Sch\"on, and A. Shnirman,
Rev. Mod. Phys. \textbf{73}, 357 (2001).
\bibitem{Rim03}
W. Lu, Z. Ji, L. Pfeifer, K.W. West, and A.J. Rimberg,
Nature \textbf{423}, 422 (2003).
\bibitem{Kor01}
A. N. Korotkov, Phys. Rev. B \textbf{63}, 085312 (2001);
{\it ibid} \textbf{63}, 115403(2001).
\bibitem{Moz04}
D. Mozyrsky, I. Martin, and M. B. Hastings,
Phys. Rev. Lett. \textbf{92}, 018303 (2004).
\bibitem{Wis06}
N. P. Oxtoby, H. M. Wiseman, and H. B. Sun,
Phys. Rev. B \textbf{74}, 045328 (2006).
\bibitem{K-A01}
A. N. Korotkov and D. V. Averin, Phys. Rev. B {\bf 64}, 165310 (2001).
\bibitem{Ave02}
D. V. Averin, Phys. Rev. Lett. \textbf{88}, 207901 (2002); A. N.
Jordan and M. B\"uttiker, Phys. Rev. B {\bf71}, 125333 (2005).
\bibitem{Wang07}
S. K. Wang, J. S. Jin, and X. Q. Li, Phys. Rev. B {\bf75},
155304(2007)
\bibitem{But06}
A. N. Jordan and M. B\"uttiker, Phys. Rev. Lett. {\bf 95}, 220401 (2005).
\bibitem{Gur05}
S. A. Gurvitz and G.P. Berman, Phys. Rev. B \textbf{72}, 073303 (2005).
\bibitem{Li051}
X. Q. Li, P. Cui, and Y. J. Yan,
Phys. Rev. Lett. {\bf 94}, 066803 (2005).
\bibitem{Li053}
J. Y. Luo, X. Q. Li, and Y. J. Yan,
Phys. Rev. B {\bf 76}, 085325 (2007).
\bibitem{Ens06}
S. Gustavsson, R. Leturcq, B. Simoviè, R. Schleser, T. Ihn,
P. Studerus, K. Ensslin, D.C. Driscoll, and A.C. Gossard,
Phys. Rev. Lett. \textbf{96}, 076605 (2006);
T. Fujisawa, T. Hayashi, R. Tomita, and Y. Hirayama,
Science \textbf{312}, 1634 (2006).
\end{references}
\end{document}